\documentclass[showpacs,twocolumn,preprintnumbers,amsmath,amssymb,aps, prb,floatfix,groupedaddress]{revtex4}
\usepackage{graphicx} 
\usepackage{dcolumn}  
\usepackage{bm}       
\usepackage{epsfig}
\usepackage{comment}
\usepackage{color}
\usepackage{multirow}
\usepackage{footmisc}

\begin{document}
\title{Analysis of charge states in the mixed valent ionic insulator AgO} 
\author{Yundi Quan and Warren E. Pickett}
\affiliation{Department of Physics, University of California Davis, Davis, CA 95616}

\date{\today}
\begin{abstract}
The doubly ionized $d^9$ copper ion provides, originally in La$_2$CuO$_4$ and later in many more
compounds, the platform for high temperature superconductivity when it is forced toward higher
levels of oxidation. The nearest chemical equivalent is Ag$^{2+}$, which is almost entirely
avoided in nature.
AgO is an illustrative example, being an unusual nonmagnetic insulating compound  with an open $4d$ shell
on one site. This compound has been interpreted in terms of one Ag$^{3+}$ ion at the 
fourfold site and one Ag$^{+}$ ion that is twofold coordinated. We analyze more aspects of 
this compound, finding that indeed the Ag$^{3+}$ ion supports only four occupied $4d$-based 
Wannier functions per spin, while Ag$^+$ supports five, yet their physical charges are nearly 
equal. The oxygen $2p$ Wannier functions display two distinct types of behavior, one type 
of which includes conspicuous Ag $4d$ tails. Calculation of the Born effective charge tensor 
shows that the mean effective charges of the Ag ions differ by about a factor of two, roughly 
consistent with the assigned formal charges. We analyze the $4d$ charge density and discuss 
it in terms of recent insights into charge states of insulating (and usually magnetic)
transition metal oxides. What might be expected in electron- and hole-doped AgO is discussed
briefly.
\end{abstract}

\maketitle

\section{Introduction}
Interest in understanding the charge (or oxidation) state of transition metal cations in oxides and halides has 
recently resurfaced,\cite{COprl,Sit2010,rappe,wepJPCM} partly because it has been established that in many charge 
ordered systems discussed as being examples of disproportionation, the actual $d$ electron occupation (i.e the
charge) is invariant, to a high degree of accuracy.\cite{COprl,Sit2010,rappe,wepJPCM} There is interest also in obtaining an objective specification 
of the charge state of an ion,\cite{Sit2010,rappe,wepJPCM}, since approaches that divide the total charge density 
amongst the various atoms have not yet proven very useful. \cite{Enatom2007,Bader}The rare earth nickelates have 
become one testing ground of models and theories, with a 2Ni$^{3+}$ $\rightarrow$ Ni$^{2+}$ + Ni$^{4+}$ 
disproportionation originally being envisioned to be responsible for the structural change and the accompanying 
metal-to-insulator transition. It has however become clear, at least for a few prominent examples,\cite{COprl,wepJPCM} that the signatures 
of ``charge order'' -- the ionic radii, the magnetic moment, and splitting of core level energies -- are obtained 
faithfully from density functional theory calculations in which there is no change in the actual charge ($4d$ occupation) 
of the Ni ion. The differences are due to the local environment: the available volume and the potential from neighboring 
oxygen ions, and the $4d-2p$ rebonding that will accompany a change in local environment.

There is interest in the doping of noble metal atoms to a valence higher than the common $1+$ configuration of
the closed shell ion, which occurs in typical band insulators. The $2+$ state of noble metal atoms is of special interest,
since the doping to higher levels for Cu, as from the La$_2$CuO$_4$ compound, leads the highest temperature superconductors (HTS) known, but with no increase in the superconducting critical temperature being achieved in the last
twenty years. The square planar environment of the Cu$^{2+}$ ion seems uniquely suited as the platform for HTS, 
but the underlying
reasons are still under debate, making study of square planar Ag (and possibly Au) of obvious interest.

The two dimensional Ag$^{2+}$ compound Cs$_2$AgF$_4$ provoked interest and study because it
is isostructural with the original cuprate superconductor
La$_2$CuO$_4$ and contains a sister doubly charged
noble metal ion,
thereby raising hopes that it might also provide excellent superconductivity when doped.
This compound, which also exists with other alkali cations, had been 
discovered\cite{hoppe, hoffmann}  before
superconductivity in the cuprates was found, and indeed the Ag ion 
is magnetic. The electronic structure is significantly
different, however, and instead of being a high temperature antiferromagnet it is a low temperature
ferromagnet -- the magnetic coupling of Ag through F is distinctly different from that
of Cu through oxygen.\cite{deepa} This compound thus became another example in the 
series of dashed hopes of finding a cuprate analog. 

The unconventional insulating and nonmagnetic compound AgO presents related but distinct questions. 
The two Ag sites Ag1 and Ag2 consist of an O-Ag1-O linear trimer and an Ag2O$_4$ unit, in which this
Ag ion is square planar coordinated (the unit is not precisely square). 
AgO has been characterized as mixed valent\cite{yvon1986,tissot1987,bielmann2002}
Ag$^{1+}$ and low-spin Ag$^{3+}$ respectively, {\it i.e.} Ag(I)Ag(III)O$_2$. It should be emphasized
that this difference is not a matter of disproportionation; the two sites are simply very different
from the moment of crystallization.
For transparency in nomenclature, in this paper we make the designation of the two Ag sites, and formal
charges, explicit by using Ag$^{III}$ and Ag$^I$ rather than Ag1 and Ag2, respectively. 
We have shown elsewhere that the two sites contain the same amount of $4d$ charge, determined by the 
same radial charge density in the vicinity of the $4d$ peak. The question is how to reconcile the 
apparent discrepancies -- differing charge states with the same d charge -- and more importantly 
how to understand differing charge states more generally in a microscopic manner.

The Ag ion is rather unusual. In the overwhelming number of compounds it is the simple closed shell ion Ag$^{1+}$ $4d^{10}$. In a few systems it behaves differently. In Ag$_4$(SO$_3$F) the Ag ion is suggested\cite{Ag4SO3F} to be mixed valent 2Ag$^{1+}$ + Ag$^{2+}$, with two different sites as in AgO, surprising because Ag almost always avoids the divalent state. When it has no choice, as in AgF$_2$ with F being the most electronegative element, the magnetic $4d^{9}$ configuration emerges, and this compound becomes a canted antiferromagnet\cite{AgF2} below T$_N$=163K. A more exotic case arises in the double perovskite Cs$_2$KAgF$_6$, which remains cubic\cite{Hoppe} because Ag acquires the high spin 3+ state which has cubic symmetry ($e_g^2$ spin-down holes).\cite{TJia} 

The most interesting analog of the cuprates arose in the Cs$_2$AgF$_4$, which is (almost) isostructural with
La$_2$CuO$_4$ and contains the same fourfold coordinated dipositive noble metal ion, and it has other
alkali metal counterparts. Chemically, it turns out, this structural analog is chemically rather different,
being for instance a ferromagnetic insulator rather than an antiferromagnetic insulator, and having an ordering 
temperature around 25 K rather than around 300 K. Thus its behavior is distinct and, so far, there is not
indication of superconductivity in this system. It may be premature however to conclude that the chapter
is closed on this cuprate analog.

In this paper, we focus on charge state characterization of the charge states of Ag$^{III}$ and 
Ag$^I$ sites in AgO by using various methods including Born effective charge, real space charge 
density and Wannier function analysis. We demonstrate for this unusual case that the charge state picture bears 
no relation to the static charge distribution, instead simply reflecting
the local environment including the distance of negative ions and the resulting $d-p$ bonding.
Nevertheless, the formal charge provides a physical description that has important consequences 
for the understanding of the physical properties of this compound.

\section{Structure and Methods}
\begin{figure}[htp]
\includegraphics[width=0.5\textwidth]{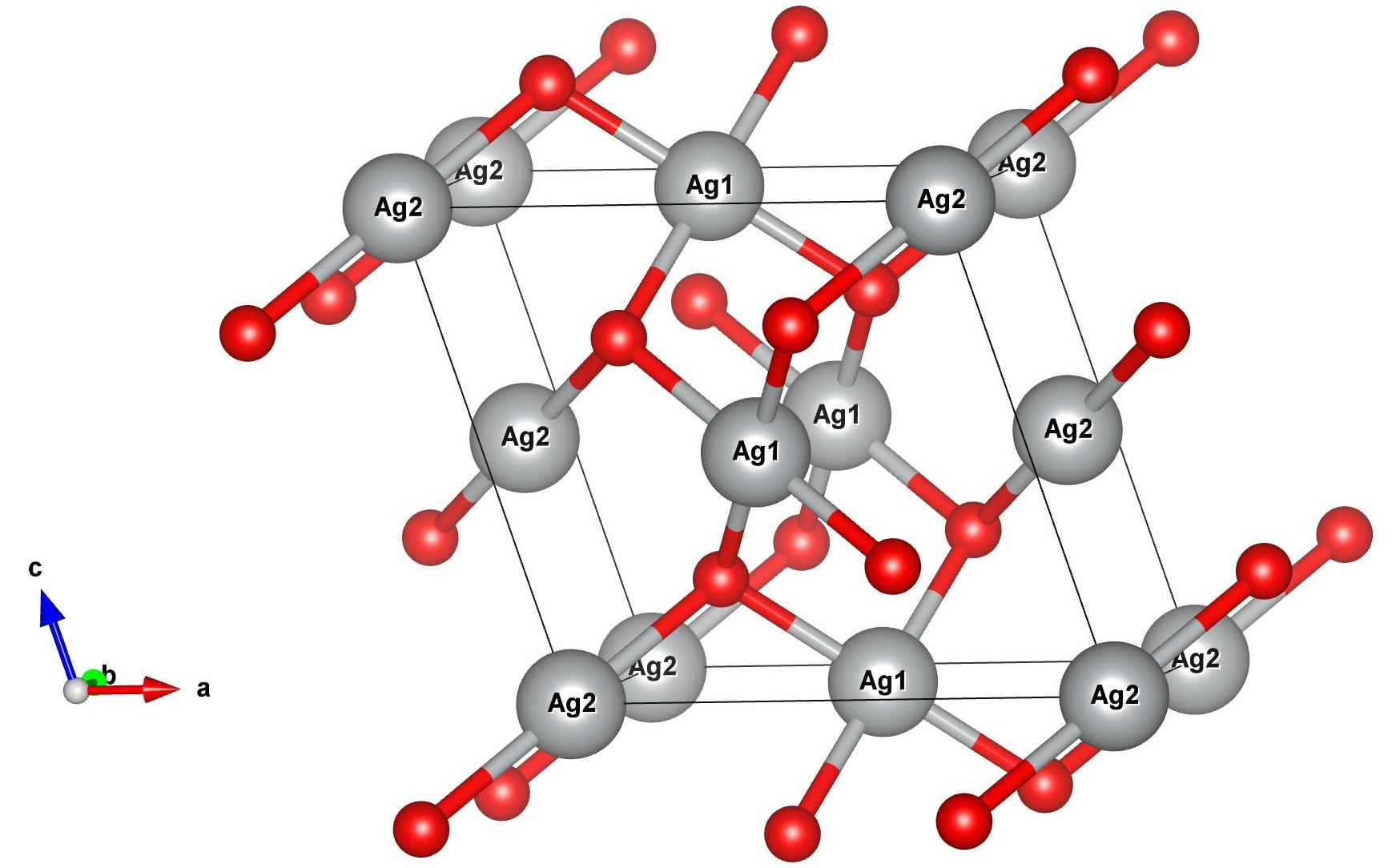}
\caption{Crystal structure\cite{momma2011} of monoclinic AgO. Ag$^{III}$ sits at the center of a 
slightly distorted but planar oxygen square, while Ag$^I$ forms an O-Ag-O linear trimer with its 
nearest neighbors. Differences in coordination numbers and bond lengths suggest distinct charge 
states for Ag$^{III}$ and Ag$^I$.}
\end{figure}

The space group of crystalline AgO is monoclinic P2$_1$/c ($\#$14), with two non-equivalent Ag sites (Ag$^{III}$, Ag$^I$) but a single oxygen site, $2d$, $2a$ and $4e$ respectively. Ag$^{III}$ sits at the center of a planar, slightly distorted oxygen square, with bond lengths 2.03~\AA \vspace{1pt} and 2.02~\AA, while Ag$^I$ and its nearest neighbor oxygens form a linear trimer with bond length 2.16~\AA. There is a single O site and each oxygen is shared by two squares and one trimer. Since the Ag$^I$-O bond is 0.13~\AA \vspace{1pt} longer than the Ag$^{III}$-O bond, and the Ag$^{III}$ coordination number is twice that of Ag$^I$, the charge state of Ag$^{III}$ and Ag$^I$ have been characterized as 3+ and 1+ to balance 2O$^{2-}$. The corresponding $4d$ occupation of Ag$^{III}$ and Ag$^I$ are (low spin) (S=0) $4d^{8}$ and closed shell $4d^{10}$ respectively, in agreement with insulating diamagnetism of AgO.

We have carried out DFT calculations using the linearized augmented plane wave method as implemented in 
WIEN2k.\cite{blaha2001} The local density approximation (LDA) exchange-correlation functional of 
Perdew and Wang\cite{PW92} was used. Because Ag$^{III}$ is an open-shell ion, we applied the LDA+U method to probe 
correlation effects beyond LDA, in spite of the low-spin (nonmagnetic) state of the ion. $U$ was 
increased from 0.1Ry with both the ``around mean field" (AMF) and ``fully localized limit" (FLL) double-counting 
functionals.\cite{Erik} In both cases, the band gap increases slowly and almost linearly with 
increasing $U$, by around 50 meV
per 1 eV increase in U, thus a U value of 4 eV (perhaps an upper limit) still leaves a severe underestimate
of the observed gap.

In recent years Wannier functions, which are localized orbitals obtained as lattice Fourier transforms of linear combinations of Bloch states, have become an indispensable tool in analyzing properties of crystalline insulators. The indeterminacy of the arbitrary phases involved in mixing the  Bloch functions, and consequently the unitary transformation matrix, leads to extra degrees of freedom which can be eliminated by imposing certain conditions (choosing gauges). In this paper, we use maximally localized Wannier functions (WFs) which are obtained by minimizing the spread functional.\cite{nicola2012,marzari1997} The overlap matrix calculation and post-processing were carried out with {\it wien2wannier}\cite{wien2wannier} code.  For the spread functional minimization, we used Wannier90,\cite{wannier90} which is independent of the basis functions used in DFT calculations.  \section{Results} \subsection{Previous work} A few theoretical studies of AgO using first principle method have been reported. Park {\it et al }\cite{park1994}  addressed the difference between Ag$^{III}$ and Ag$^{I}$ by analyzing partial densities of states (PDOS). They  demonstrated that Ag$^{I}$ has significant spectral weight and strong hybridization with oxygen only in the region below the Fermi level, while there exists a Ag$^{III}$ d peak immediately above the Fermi level which is equally strongly hybridized with oxygen. At the experimental geometry, LDA and GGA give an almost exactly vanishing
gap (using a 5000 point k-mesh for self-consistency). 
Recent work by Allen {\it et al.} using pseudopotential methods with a hybrid functional 
(part Hartree-Fock exchange, part local density exchange) calculation gives a direct band gap of 
1.2eV\cite{jeremy2010,jeremy2011} which is consistent with the optical band gap of 1-1.1eV observed from 
experiment.\cite{bandgap} 

Further discussion of the $4d-2p$ hybridized contour and PDOS was provided by
Pickett {\it et al.},\cite{wepJPCM} including a plot of the charge density displaying all ions from the unoccupied
``Ag$^{III}$O$_4$" band, which contains significant Ag$^I$ character as well. These authors also noted that there
is negligible difference in $4d$ occupation on the two Ag sites -- the spherically averaged charge densities in the
vicinity of the $4d$ peak are the same -- and that the orbital occupations of the two sites strains the rule
proposed by Sit {\it et al.}\cite{Sit2010} that formal charges of ions can be obtained from orbital 
occupations that are used in LDA+U
calculations.   

\begin{figure}[htp]
\includegraphics[width=0.5\textwidth]{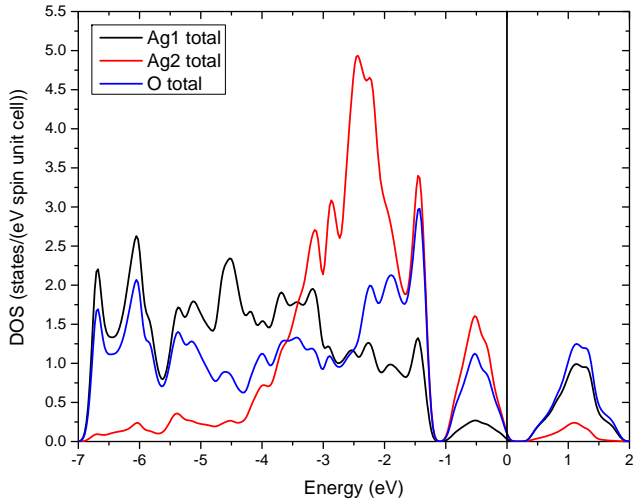}
\caption{Atom-projected density of states of AgO resulting from the application of the TB-mBJ
potential, as described in the text. The bottom of the gap
is taken as the zero of energy.} 
\label{PDOS}
\end{figure}

For orientation the atom-projected density of states (PDOS) is presented in Fig.~\ref{PDOS}. These results
are not for LDA or GGA, however, which as mentioned above give a very small gap. These PDOSs are the
result of applying the modified Becke-Johnson (mBJ) potential, which is known to greatly improve on
the gaps of a few classes of insulators, compared to the LDA or GGA result. The form we have used is the
Tran and Blaha modification\cite{TranBlaha,mBJ} TB-mBJ. It incorporates the orbital kinetic
energy density into the potential as a improved way\cite{SharpHorton,Talman} 
of representing the non-local exchange potential,
with little expense beyond reconverging to self-consistency. For AgO, the mBJ potential leads to a gap of 0.4 eV.

Fig.~\ref{PDOS} illustrates the very different spectral distributions of the Ag1 (Ag$^{III}$) and
Ag2 (Ag$^I$) $4d$ states.  The former has weight distributed throughout the -6.5 eV to -1 eV region,
whereas the latter has its weight concentrated in only the upper half of this range -- its states are
much less strongly bound. In the 1 eV range below the valence band maximum, there is one band per Ag pair per spin that has a majority of Ag$^{I}$ weight, however it has substantial Ag${III}$ weight as well. Above the gap is one more band per Ag pair per spin that is heavily Ag$^{III}$ hybridized with O $2p$, and very little Ag$^{I}$ weight; this band has been identified as
the unoccupied band of the $4d^*$ ion.

\begin{figure*}[!ht]
\includegraphics[width=0.99\textwidth]{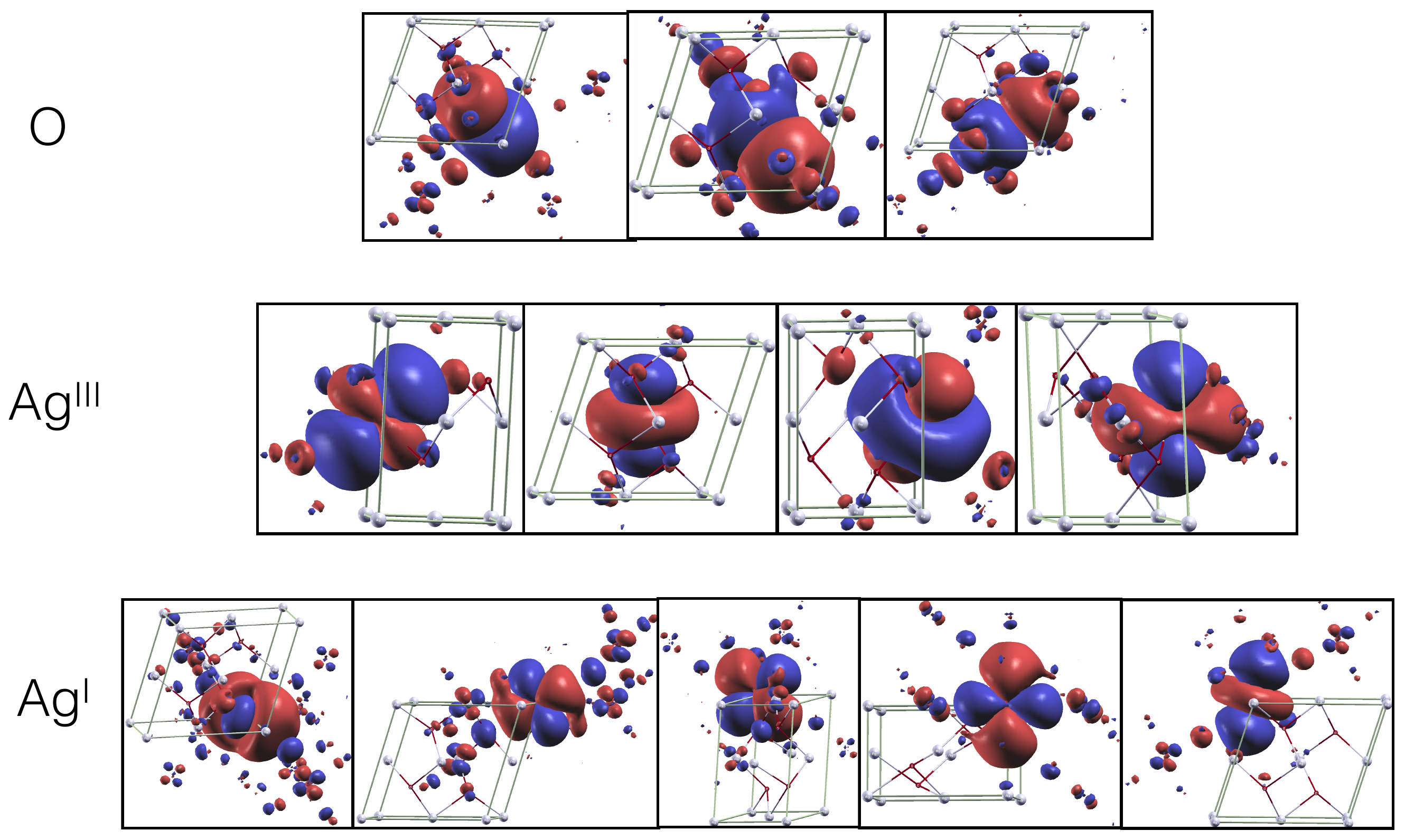}
\caption{Isosurface plots of the Ag $4d$ maximally localized
Wannier functions of AgO, red and blue denote opposite signs.
The top three are O-centered, $2p$-based; note that the latter two
have substantial contributions from an Ag orbital only at one end.
The next four are centered at Ag$^{III}$ $d^8$ site and 
 are more confined but larger on the Ag site, while the bottom five
are centered at the Ag$^I$ $d^{10}$ site. Each of the four Ag$^{III}$ WFs
contains more $4d$ charge than does an Ag$^I$ WF, but the latter
WFs include more neighboring O $2p$ contribution.
All isosurface values are the same. These WFs were obtained from a 
10$\times$16$\times$11 Brillouin zone mesh.}
\label{wfs}
\end{figure*}

\subsection{Wannier function analysis of AgO}

The formal charge states correspond to four occupied orbitals per spin for Ag$^{III}$, five for Ag$^I$, and of course three for each O$^{2-}$. The manifold of bands used for Ag$^{I}$Ag$^{III}$O$_2$ in the WF calculation consists of 30 isolated valence bands immediately below the gap, which are Ag $4d$ bands mixed with oxygen $2p$ bands. The spread functional minimization leads to localized site-centered $4d$-derived WFs at Ag$^{III}$ and Ag$^I$ sites, and Wannier functions at the O sites that, notably, are not exactly site centered. 

Isosurfaces of the WFs for AgO are plotted in Figs 3. We obtain four WFs centered at Ag$^{III}$ site and five centered on Ag$^I$. The WFs at the Ag$^I$ site, shown in Fig. 3, have at their core the classic shape of the five $d$ orbitals. Their orientation is non-intuitive, moreover their extension includes neighboring O $2p$ contributions
and even neighboring Ag $4d$ contributions. The four WFs at the Ag$^{III}$ site have a core that is of $4d$ shape as well. However, one of the five $4d$ orbitals is missing; on the other hand, the volumes at the chosen isosurface are
larger than for the Ag$^I$ WFs, that is, the unit charge of these WFs contain more contribution from the $4d$
orbitals than do the WFs centered on Ag$^I$.

While the WFs of Ag are centered on the nuclear sites, the centers of the O WFs differ from oxygen nuclear positions, 
by values of 0.13~\AA and 0.33~\AA (twice). These displacements reflect first, polarization of the O$^{2-}$ ions that is non-intuitive due to its low site symmetry, and secondly, the substantial admixture of Ag contributions to two of them.   In the figure, it can be seen that of the three O $2p$ WFs, one has exemplary O $2p$ character while the other two represent O $2p$ orbitals that are strongly mixed with an Ag$^{III}$ $4d$ orbital - one $4d$ tail extends from each of two O $2p$ WFs. The mixing of Ag$^{III}$  $4d$ orbital to the oxygen $2p$ WF is aligned along the diagonal of the oxygen plaquette. Consequently, the Wannier functions  for oxygens sitting at opposite ends of the diagonal have significant overlap.

\begin{figure}[htp]
\includegraphics[width=0.45\textwidth]{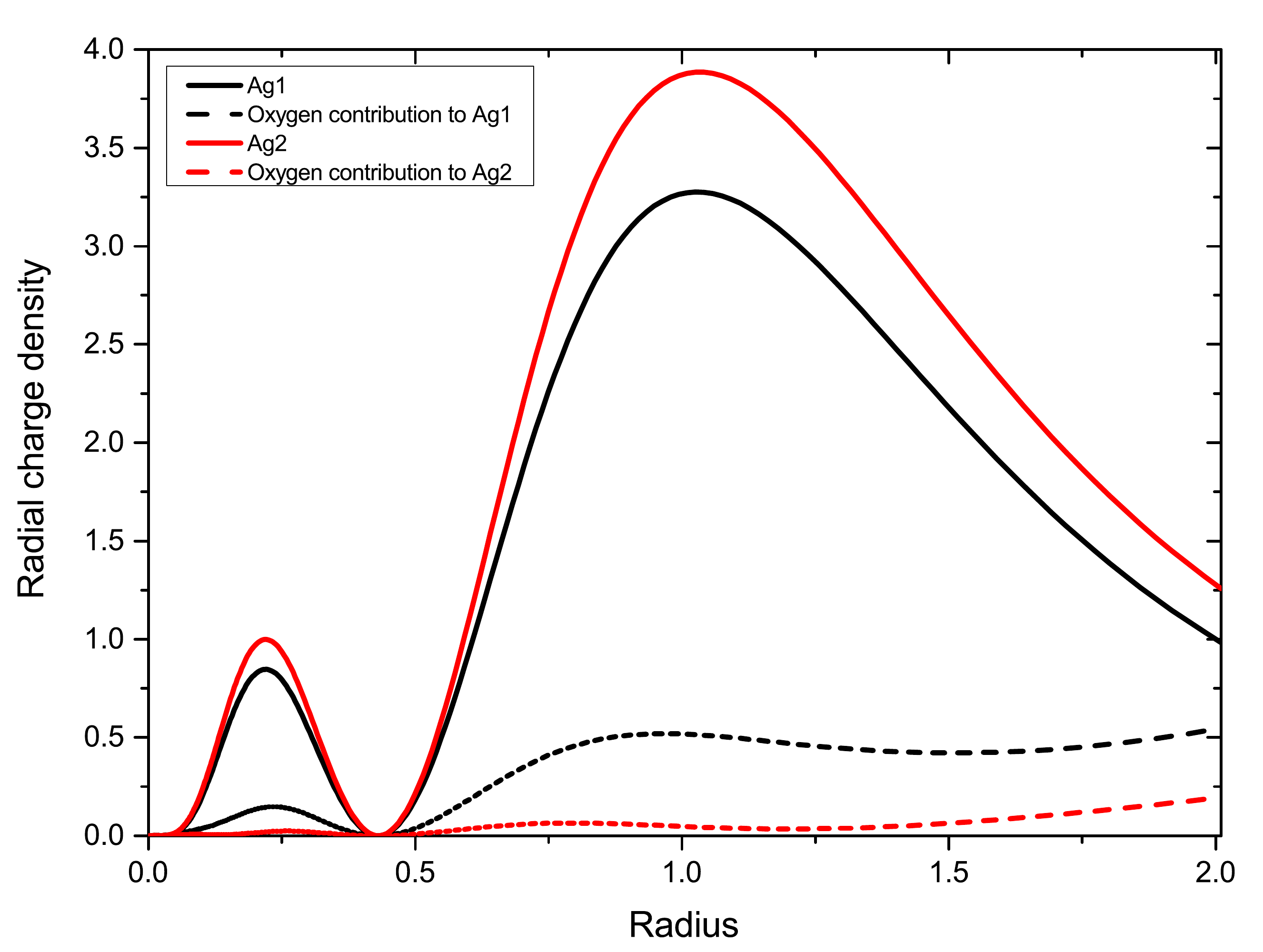}
\caption{Black solid line: radial charge densities 4$\pi r^2 \rho(r)$ versus distance from the Ag nucleus, in a.u.  
Red solid line: sum from the five Ag$^I$ WFs. Black solid line: sum from four Ag$^{III}$ WFs. Dashed lines: corresponding 
contributions from the O $2p$ WFs within the Ag spheres.The O WFs make large contributions to the Ag$^{III}$ atomic density.}
\end{figure}

To  assess the charge distributions, the decomposed radial charge densities inside 
Ag$^{III}$ and Ag$^I$ spheres are displayed in Fig. 4. The black and red lines are 
the decomposed radial charge densities of WFs in the Ag$^{III}$ 
and Ag$^I$ spheres respectively. Solid lines are the contribution from $4d$ WFs 
(and their translated images); dashed 
lines are the contribution from oxygen-based WFs. 
The lower contribution to the $4d$ occupation from four WFs of Ag$^{III}$, 
versus five for Ag$^I$, are almost exactly compensated by the contribution from the 
two of the O-centered WFs. 
The resulting equivalence of real charge on ions characterized by different charge 
states has become a recurring theme 
in magnetic ``charge-ordered" compounds, thus comes over to nonmagnetic AgO.\cite{COprl,wepJPCM}

\subsection{Comments on energies}
WFs provide an orthonormal tight-binding representation of the band structure in the region spanned by the WFs. 
Characteristics of interest include the on-site energies and hopping amplitudes, 
the latter of which for nine Ag WFs and six O WFs are
too numerous to present. For these maximally localized WFs, we find the interactions are mostly short-range. The band 
structure obtained from diagonalizing the Hamiltonian with truncated summation (after few neighboring cells) agrees 
well with the band structure from DFT calculation. 

The onsite energies of all occupied WFs are listed in Table I. Ag$^{III}$ on-site energies range from -4.49eV to -3.94eV, with one split off from the rest by 0.5 eV. The Ag$^I$ on-site energies range from -2.86eV to -2.59eV, a rather small range for transition metal ions in  oxides. The oxygen on-site energies are more interesting, ranging over nearly 1 eV, with a pair at -4.30$\pm$0.03 eV and the other at -3.42eV. As mentioned above, two of the oxygen WFs have substantial Ag $4d$ tails inside Ag$^{III}$ sphere, and the on-site energies reflect this distinction. Hopping parameters between oxygen Wannier functions that extend into a given Ag$^{III}$ site are as large as -1.0 eV, reflecting the strong interaction between neighboring 
O orbitals through Ag$^{III}$ orbitals. 

The fifth Ag$^{III}$ WF would correspond to the unoccupied band just above $E_F$. Due to the band crossing with a
higher lying band, there is entanglement with conduction band WFs and we have not obtained this WF. Its
site energy presumably would lie much nearer to, if not above, the bottom of the gap.  Note
also that there is not a distinct Ag$^I$-derived WF corresponding to the $4d$ band 0.5 eV below E$_F$;
all of the Ag$^I$ WFs have site energies around -2.7 eV. 

\begin{table}[!ht]
\begin{tabular}{c|ccccc|c}
\hline
Atom & \multicolumn{5}{c|}{on-site energies } & mean\\
\hline
Ag$^{III}$ & -3.94 & -4.49 & -4.44 & -4.31 & --- & -4.29\\
\hline
Ag$^I$ & -2.59 & -2.73 & -2.82 & -2.86 & -2.78  & -2.76\\
\hline
O & -3.42 & -4.27 & -4.33 & --- & --- &   -4.00        \\
\hline
\end{tabular}
\caption{On-site energies (eV) of the occupied Wannier functions described in the text and shown in Fig. \ref{wfs}. 
Note that for Ag$^{III}$ and for O one site energy is separated substantially from the others, while there 
is little crystal field splitting for Ag$^{I}$. Since the GGA gap is vanishingly small
(slightly negative), these site energies
were checked up to a 14$\times$22$\times$15 Brillouin zone integration mesh.}
\end{table}

\subsection{Born Effective Charge of AgO}
The Born effective charge (BEC) tensor $Z_{\alpha\beta}$ is a measure of the dynamic response of electron system to ionic displacements. Vibrating ions drive electric response as if they had this charge. Using the finite  displacement method, we calculated the BECs of Ag$^{III}$, Ag$^I$ and O in Cartesian coordinates.  The displacements $\Delta x_j$ along $\vec{a}$, $\vec{b}$ and $\vec{c}$ were 0.001 in internal units, within the linear response regime. Since the lattice vectors of AgO are non-orthogonal, we chose to have $\hat{y}$ and $\hat{z}$ pointing along $\vec{b}$ and $\vec{c}$ respectively. In the Cartesian coordinate system, the change in polarization $\Delta P$ is related to displacements $\Delta x_j$ by 
\begin{equation}
\Delta P_i =  \frac{e}{\Omega}  \sum_{\substack{j=x,y,z}} Z_{ji}\Delta x_j \; \; \; \; \; (i=x,y,z)
\end{equation}
where $\Omega$ is the cell volume.
Eighteen separate calculations were carried out displacing Ag$^{III}$, Ag$^I$ and O along $\vec{a}$, 
$\vec{b}$ and $\vec{c}$ respectively. Polarization was calculated using the BerryPi code.\cite{BerryPi} 
We used the LSDA+U method with U=4.08 eV, J=0.68 eV. 
We obtained the Born effective charge tensors for Ag$^{III}$, Ag$^I$ and O, shown in Table 2.

\begin{table}
\begin{tabular}{ccc|ccc|ccc}
\hline
\multicolumn{3}{c|}{Ag$^{III}$}& \multicolumn{3}{c|}{Ag$^{I}$} & \multicolumn{3}{c}{O}\\
\hline
 0.83 & -0.96 &  0.58 &  1.49 &  0.34 & 0.06 & -1.15 & 0.10 & -0.33 \\
-0.20 &  2.16 & -1.20 &  0.32 &  0.72 & 0.49 & 0.15 & -1.43 & 1.11 \\
-0.27 &  0.34 &  3.58 & -0.14 & -0.16 & 1.12 & 0.19 & 0.82 & -2.35 \\
\hline
\multicolumn{3}{c|}{2.94 $\pm$ $0.41i$ 0.69 } & 1.51 & 1.13 & 0.69  & -2.93 & -1.11 & -0.89 \\
\hline
\end{tabular}
\caption{The Born effective charge tensors for Ag$^{III}$, Ag$^I$ and O. The eigenvalues are
provided on the bottom line.}
\end{table}

The unusual open shell but low spins Ag$^{III}$ ion has complex eigenvalues of the BEC tensor. 
We confirm that $\sum_{\substack s} Z_{s, \alpha \alpha}=0$ ($\alpha=x,y,z$) as required by the 
acoustic sum rule. Eigenvalues (see Table II) of $Z_{Ag^I}$ are in rough agreement with a 1+ 
formal charge state characterization. Eigenvalues of $Z_{O}$, -2.93, -1.11 and -0.89, are highly 
direction dependent. The modulus of the complex eigenvalues (complex conjugates) of Ag$^{III}$ 
is 2.97, which is close to its formal valence. 

\section{Comments on doping}
Extrapolating from the doping of Cu$^{2+}$ states that are square coordinated with oxygen
and lead to high temperature superconductivity, possible effects of doping of AgO become
of interest. Typical samples of AgO are slightly doped $n$-type,\cite{bielmann2002} such as
might arise from oxygen vacancies, and thin Ag-O films that
are transparent conductors are believed to involve AgO. The 1 eV bandwidths of both the 
highest occupied band and the lowest unoccupied band (see Fig.~\ref{PDOS}), together with the nearly 100\%
underestimate of the gap by the local density approximation, suggests that correlation
effects will be important for the doped carriers, which might arise from either Coulomb
interactions or electron-lattice interaction. 

A guide to what can be expected
can be obtained from the PDOS shown in Fig.~\ref{PDOS}. Excess electrons will 
inhabit the empty Ag$^{III}$
band, where mixing with O $2p$ is very strong. The relevant orbital is the 
corresponding Ag$^{III}$O$_4$
cluster orbital (Wannier function). 
Strong electron-lattice coupling may result in polaronic
conduction or, at low doping, localization. The carriers are likely to be magnetic on a time scale that
could lead to a large susceptibility per carrier.

The small bandwidth and strong interaction aspects should apply to hole doping as well.
A qualitative difference is that holes enter $4d$ states on the two-fold coordinated Ag$^I$ ion.
Hopping to neighboring Ag$^I$ sites are strongly inhibited however, because each O ion
is linked to two square units but only a single trimer. A hole on a trimer must
tunnel through a square to reach another trimer, in order to move through the lattice.
Thus there will be a much stronger tendency for holes to localize and also to be magnetic,
than for electrons. This picture is consistent with conductivity in AgO being $n$-type. 

Much the same behavior can be seen in lightly hole-doped cuprates, where antiferromagnetic
order (or strong correlations) also contribute. Approximately 5\% holes must be doped in,
only then does the system become conducting and superconducting. There are of course
several differences besides the magnetism between the AgO compound studied here and the
cuprates: two dimensional layers, relatively straightforward hopping of carriers, and
several others. The similarities do seem intriguing enough that doping studies of AgO should be
pursued.

\section{Summary}
In this paper we have studied the electronic structure, the ionic character, and the dynamic linear response of the system to ionic displacements.  Specifically, we have analyzed the character of the two very distinct Ag sites focusing on a ``charge state'' (formal oxidation state) picture. As we have found in several related ``charge disproportionated'' materials,\cite{wepJPCM} the $4d$ occupations of the two Ag ions are nearly the same, based on the magnitude of the Ag radial charge density in the vicinity of the $4d$ peak. On the other hand, maximally localized Wannier functions are calculated, finding that there are four $4d$-like Wannier functions per spin centered on Ag$^{III}$ while there are five $4d$-like Wannier functions per spin on Ag$^{I}$. This distinction, as well as the computed Born effective charges, are in agreement with the 3+ and 1+ characterization of the charge states. 

The characters of the WFs at the two sites are different: the Ag$^{I}$ WFs include larger contributions from neighboring oxygens, with noticeable contribution from neighboring Ag sites as well.  Additionally, the average on-site energies differ by 1.5 eV. The three O WFs also display two types of behavior: two have strong admixtures with Ag$^{III}$ $d$ orbitals, while the other is of the more classic type of nearly pure oxygen $2p$ orbital. The two similar WFs differ in site energy by almost 1 eV from the third. Only when contributions from all Ag and O WFs are included do the charge densities on sites Ag$^{III}$ and Ag$^I$ become nearly the same. 

The Born effective charge tensors reflect the low site symmetries of the ions, and reveal very strong directional dependence. For Ag$^{III}$, the maximum diagonal element is 3.6, very consistent with a 3+ designation.  The minimum element is however only 23\% of that. For Ag$^I$, the values range over a factor of two.  Still, the rms values of the moduli of the tensor eigenvalues differ by a factor of two, indicating large differences consistent with very different formal valences.  Overall, our calculations support (1) the longstanding view that the two Ag sites merit 1+ and 3+ designations, and (2) our recent suggestion\cite{wepJPCM} that it is the number of occupied Wannier functions  that correlates with its charge state. This work has revealed that the differences in the character of the Wannier functions for the two charge states can be substantial.

\section{Acknowledgments}  The authors would like to thank Elias Assmann for providing useful 
discussion about the wien2wannier code,  Oleg Rubel for providing BerryPi code, and many useful discussions with A. S. Botana. 
This work was supported by National Science Foundation Grant DMR-1207622.

\end{document}